\documentclass[aps,groupedaddress,superscriptaddress,amssymb,amsmath,amsbsy,amsfonts,twocolumn,pra]{revtex4}
\usepackage{amsmath}
\usepackage{amssymb}
\usepackage{graphicx,color}
\usepackage{amsthm}
\usepackage{hyperref}
\usepackage{extarrows}
\usepackage{adjustbox,lipsum}
\usepackage{makecell}
\usepackage{bbold}

\begin{document}

\newcommand*{\cl}[1]{{\mathcal{#1}}}
\newcommand*{\bb}[1]{{\mathbb{#1}}}
\newcommand{\ket}[1]{|#1\rangle}
\newcommand{\bra}[1]{\langle#1|}
\newcommand{\inn}[2]{\langle#1|#2\rangle}
\newcommand{\proj}[2]{| #1 \rangle\!\langle #2 |}
\newcommand*{\tn}[1]{{\textnormal{#1}}}
\newcommand*{\1}{{\mathbb{1}}}
\newcommand{\T}{\mbox{$\textnormal{Tr}$}}
\newcommand*{\todo}[1]{\textcolor[rgb]{0.99,0.1,0.3}{#1}}

\theoremstyle{plain}
\newtheorem{prop}{Proposition}
\newtheorem{proposition}{Proposition}
\newtheorem{theorem}{Theorem}
\newtheorem{lemma}[theorem]{Lemma}
\newtheorem{remark}{Remark}

\theoremstyle{definition}
\newtheorem{definition}{Definition}

\title{Decomposed entropy and estimation of output power in deformed microcavity lasers}
\author{Kyu-Won Park}
%\email{wind999@snu.ac.kr}
\affiliation{Research Institute of Mathematics, Seoul National University, Seoul 08826, Korea}
\author{Kwon-Wook Son}
%\email{kwson@kopti.re.kr}
\affiliation{Department of Electrical Engineering, Yeungnam University, Gyeongsan 38541, Korea}
\author{Chang-Hyun Ju}
%\email{ele0853@naver.com}
\affiliation{Department of Electrical Engineering, Yeungnam University, Gyeongsan 38541, Korea}
\author{Kabgyun Jeong}
\email{kgjeong6@snu.ac.kr}
\affiliation{Research Institute of Mathematics, Seoul National University, Seoul 08826, Korea}
\affiliation{School of Computational Sciences, Korea Institute for Advanced Study, Seoul 02455, Korea}

\date{\today}

\begin{abstract}
Park \emph{et al}. [Phys. Rev. A \textbf{106}, L031504 (2022)] showed that the Shannon entropy of the probability distribution of a single random variable for far-field profiles (FFPs) in deformed microcavity lasers can efficiently measure the directionality of deformed microcavity lasers. In this study, we instead consider two random variables of FFPs with joint probability distributions and introduce the decomposed (Shannon) entropy for the peak intensities of directional emissions. This provides a new foundation such that the decomposed entropy can estimate the degree of the output power at given FFPs without any further information.
\end{abstract}

\maketitle

%%%%%
\section{Introduction}
Deformed microcavity lasers~\cite{JA97,c96} have been widely studied up to now owing to the existence of models for investigating various physical phenomena, such as scar~\cite{E84,SS21}, tunneling~\cite{SA10,FR19}, exceptional point~\cite{KJ20,AD21}, and ray-way correspondence~\cite{SM09,HK02}. They have also attracted considerable attention owing to their potential for optical applications~\cite{v03}. In particular, because directional emissions are decisive in achieving high-performance lasers, various types of microcavity lasers~\cite{wm08,CQ09,WQ09,QL10,XC16,LZ22,ZF17} have been considered. Moreover, numerous physical systems, such as multi-layer thin films~\cite{HC22}, quantum metasurfaces~\cite{DA22}, and nanoantennas~\cite{MA17} have been studied on this issue.
In these studies, directional emission was only addressed in terms of a single variable. That is, the directionality and divergence angle are defined in terms of the intensity of far-field profiles (FFPs) as a function of a single variable $\theta$~\cite{JJ19,JM11,JA13,QL10}. However, these physical quantities cannot fully capture the properties of FFPs because the FFPs are defined in two-dimensional microcavity lasers, rather than in one-dimensional lasers.

In particular, our previous work presented the Shannon entropy to measure directionality more efficiently~\cite{KC22}. However, the Shannon entropy introduced in our work was only defined by the probability distribution of a single (random) variable, as in previous studies.
Accordingly, in this study, we consider Shannon entropy, which is defined by the joint probability distribution of two random variables in a mathematically rigorous manner. In doing so, we can further introduce the sub-sample space associated with the peak intensities of the directional emissions. We present a new quantity called \emph{decomposed entropy} in two-dimensional deformed microcavity lasers. This quantity behaves opposite to the previous Shannon entropy~\cite{KC22}. Moreover, it can effectively capture the properties of directional emission more sensitively for two different microcavity lasers. We argue that this decomposed entropy can estimate the degree of output power related to the intensities of peaks for directional emissions at given FFPs, although previous notions can only estimate the directionality and divergence angle at given FFPs.

The remainder of this paper is organized as follows. In Sec.~\ref{sec:decompent}, we briefly recapitulate the probability theory in the context of mathematics and introduce decomposed entropy. We describe our proposed system in Sec.~\ref{sec:system}. An analysis of the lima\c{c}on-shaped microcavity is presented in Sec.~\ref{sec:limacon}. In Sec.~\ref{sec:oval}, the oval-shaped microcavity is analyzed. Finally, we summarize our work in Sec.~\ref{sec:conclusion}.

%%%%%
\section{Probability space and decomposed entropy for deformed microcavity lasers}\label{sec:decompent}
%%%%%
\subsection{Recapitulation of the probability space and random variables}
 A measurable space $(S,\mathcal{A})$ with measure $\mu$ is the measure space. Here, $S$ is a non-empty set, $\mathcal{A}$ is sigma algebra, and $\mu$ is a map from $\mathcal{A}$ to a non-negative number $\mu$: $\mathcal{A}$ $\mapsto [0,\infty]$. A measurable map $f$ is a map between two measurable spaces $(S,\mathcal{A})$ and $(\tilde{S},\mathcal{\tilde{A}})$ if, for all $a\in\mathcal{\tilde{A}}$, the pre-image of $a$ under the map $f$ is the element of the algebra $\mathcal{A}$: $f^{-1}(a)\in\mathcal{A}$ for all $a\in\mathcal{\tilde{A}}$. When fulfilling the $\mu(S)=1$, the measure space $(S,\mathcal{A},\mu)$ corresponds to the probability space $(\Omega, E, \mathbb{P})$ where $\Omega$ is called a sample space, $E$ is called an event space, and $\mathbb{P}$ is a probability function that assigns a probability to each event.

A random variable typically denoted by $X$ is a measurable map from $\Omega$ to another sample space $\tilde{\Omega}$. Here, the space $\tilde{\Omega}$ typically indicates real numbers. Using these definitions, we can now define a probability mass function $\rho_{X}(x_{i})$, where we simply use the notation $\rho (x_{i})$. The probability distribution of a discrete random variable $X$ from $\mathbb{R}$ to $[0,1]$ is defined by $\rho_{X}(x_{i})=\mathbb{P}(X=x_{i})$ satisfying the normalization condition $\sum_{x_{i}}\rho_{X}(x_{i})=1$. That is, $\rho_{X}(x_{i})$ is equivalent to the probability of the occurrence of event $e\in E$ in sample space $\Omega:\mathbb{P}(X=x_{i})=\mathbb{P}\{e: X(e)=x_{i}\}$.

 To perform numerical calculations for the intensities of FFPs in microcavity lasers, we consider only discrete random variables, rather than continuous ones, in our study. In addition, we should consider the two random variables $X,Y$ to introduce the joint probability distributions for two-dimensional microcavity lasers; that is,  $\rho_{X,Y}(x_{i},y_{j})=\mathbb{P}(X=x_{i},Y=y_{j})$ under the normalization condition $\sum_{i,j}\rho_{X,Y}(x_{i},y_{j})=\sum_{i,j}\rho(x_{i},y_{j})=1$. In this case, random variables $X$ and $Y$ are the maps from the sample space (i.e., positions in $X$- and $Y$-axes) to the real numbers (i.e., components of the coordinates of $X$- and $Y$-axes). This fact imposes the probability $\rho$ at each Cartesian coordinate $(X=x_{i},Y=y_{j})$.

%%%%%
\subsection{Decomposed entropy of the peak intensities for FFPs in two dimensional microcavity lasers} \label{subsec:decomp}
Shannon information entropy is a relevant measure of the average information content or can be interpreted as the complexity of a given probability distribution of a random variable. It was first developed by Claude Shannon in his seminal paper on ``communication theory''~\cite{C48} and has been extensively exploited in numerous areas. Shannon entropy has been employed for variability, for example, in molecular descriptors~\cite{JF00} and protein sequences~\cite{BT96}. This has also been exploited in economics~\cite{J22} and market efficiency~\cite{AP22}.

The discrete Shannon entropy of probability distributions $\{\rho(x_{i})\}_{i=1}^N$ for the random variable $X$ is formally defined as
\begin{align}
H=-\sum^{N}_{i=1}\rho({x_{i}})\log \rho({x_{i}})
\end{align}
under the normalization condition $\sum^{N}_{i=1}\rho(x_{i})=1$.

First, to handle Shannon entropy of peak emissions for directional emissions, we assume that the total sample space $\Omega$ can be decomposed into sub-sample space $\bar{\Omega}$ and its complement $\bar{\Omega}^{c}$: Another words, $\Omega=\bar{\Omega} \cup \bar{\Omega}^{c}$ under the condition $\{\rho({x_{1}})$,\ldots,$\rho({x_{j}})\}\in \bar{\Omega}$ and $\{\rho({x_{j+1}})$,\ldots,$\rho({x_{N}})\}\in \bar{\Omega}^{c}$, resulting in $\sum^{j}_{i=1}\rho(x_{i}) + \sum^{N}_{i=j+1}\rho(x_{i}) =1$. Under these conditions, we can assume that the Shannon entropy can be decomposed into two parts in analogy to Boltzmann--Gibbs entropy~\cite{EC91,AA97} as follows:
\begin{align}
H_{T}=\bar{H}+\bar{H}^{c}.
\end{align}

Here, $\bar{H}=-\sum^{j}_{i=1}\rho({x_{i}})\log \rho({x_{i}})$ is defined only in $\bar{\Omega}$, and $\bar{H}^{c}=-\sum^{N}_{i=j+1}\rho({x_{i=j+1}})\log \rho({x_{i=j+1}})$ is defined solely in $\bar{\Omega}^{c}$. Employing these conventions, we analyze new properties of the decomposed entropy on two shapes (i.e., lima\c{c}on- and oval-shaped) of microcavity lasers.

%%%%%
\section{Illustration of the proposed microcavity lasers }\label{sec:system}
n our previous paper, we considered the lima\c{c}on-shaped microcavity and oval-shaped microcavity lasers as candidates for directional emissions. The geometrical boundary of the lima\c{c}on-shaped microcavity is defined as $R(\theta)=(1+\chi\cos\theta)$, and that of the oval-shaped microcavity lasers is defined as $x^2/a^2+(1+\varepsilon x)y^2/b^2=1$. For the lima\c{c}on-shaped microcavity, $\chi$ is the deformation parameter in the range of $0.43\leq\chi\leq0.478$, and $\theta$ is the angle in polar coordinates. For the oval-shaped microcavity laser, parameters $a$ and $b$ are the major and minor axes of an ellipse with fixed values $a=1.0$ and $b=1.03$, and $\varepsilon$ is the deformation parameter in the range of $0.043\leq \varepsilon\leq0.058$. Both microcavity lasers have the same effective refractive index, $n=3.3$ for InGaAsP semiconductors.

\begin{figure}%[ht!]
\centering
\includegraphics[width=\linewidth]{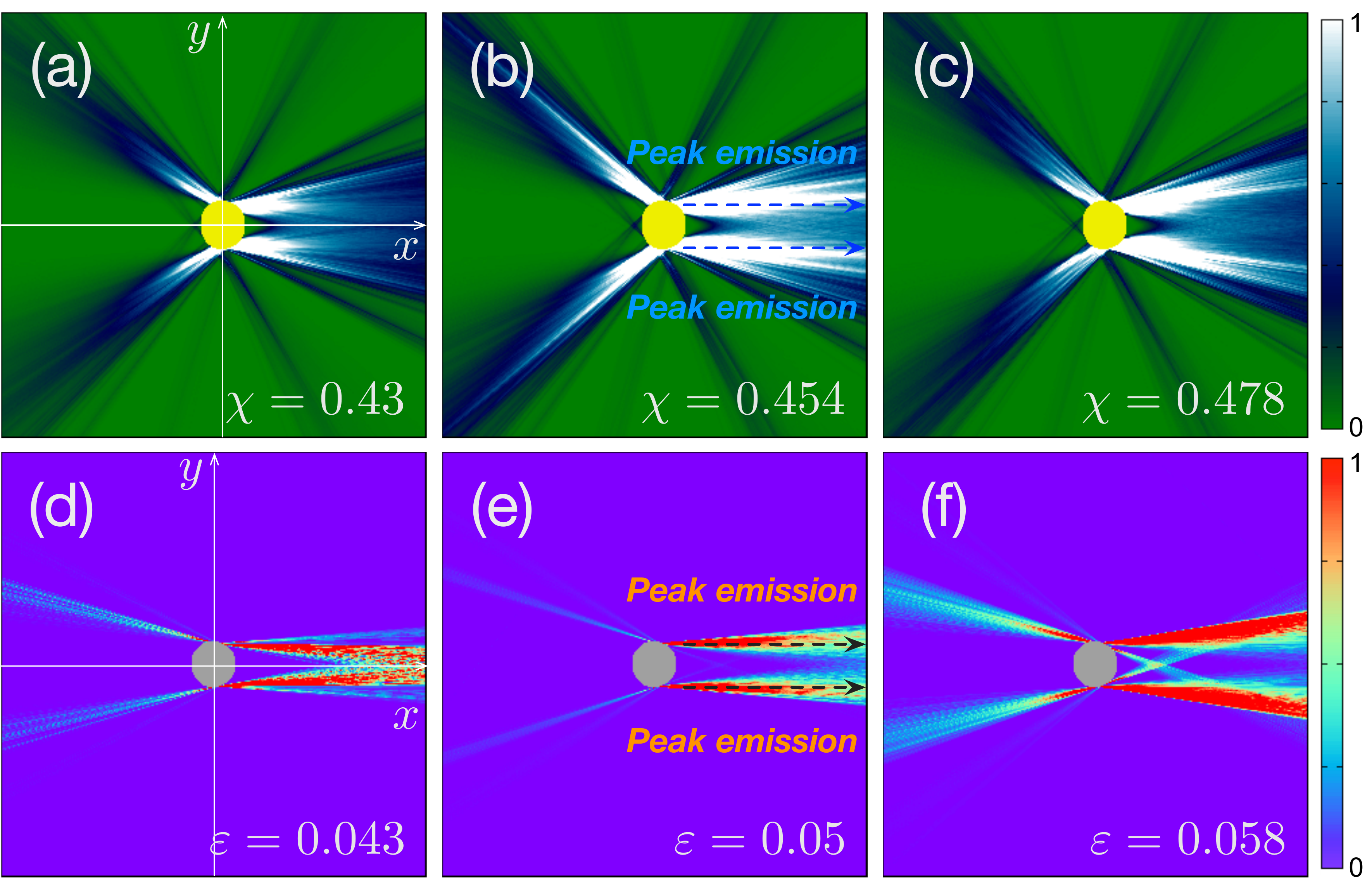}
\caption{Some of the representative far-field profiles (FFPs) in a lima\c{c}on-shaped cavity and oval-shaped microcavity lasers~\cite{KC22}. Figures (a), (b), and (c) are the FFPs of a lima\c{c}on-shaped cavity in the Cartesian coordinate within $-10\leq x\leq10$ and $-10\leq y\leq10$ at each deformation $\chi=0.43$, $\chi=0.454$, and $\chi=0.478$, respectively. Figures (d), (e), and (f) are the FFPs of an oval-shaped cavity in the Cartesian coordinate within $-10\leq x\leq10$ and $-10\leq y\leq10$ at each deformation $\varepsilon=0.043$, $\varepsilon=0.05$, and $\varepsilon=0.058$, respectively. The thick-dotted lines in Figures (b) and (e) depict two peak emissions of the deformed microcavity lasers.}
\label{fig:1}
\end{figure}

%%%%%
\begin{figure*}%[ht!]
\centering
\includegraphics[width=\linewidth]{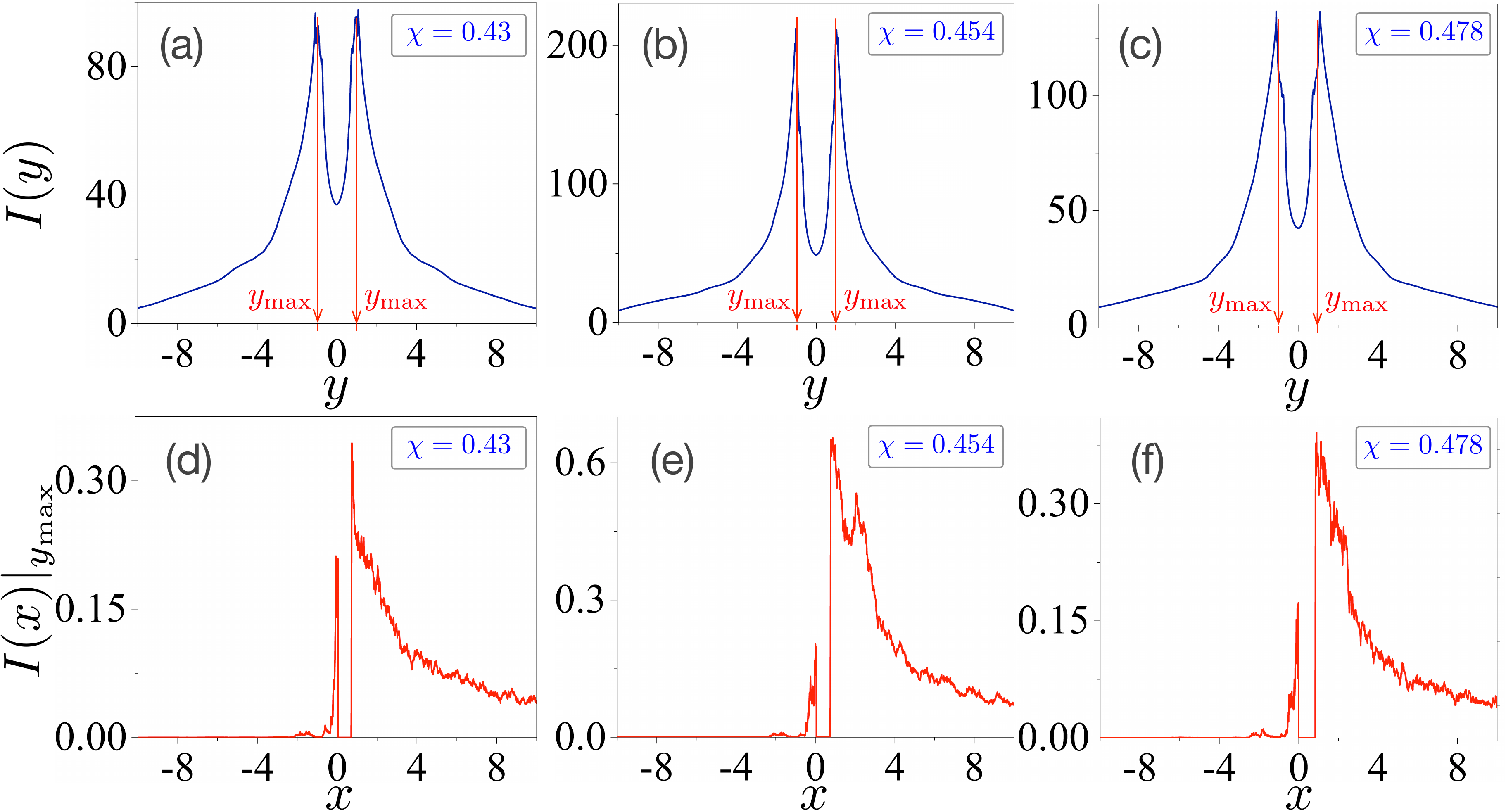}
\caption {Marginal and conditional intensities in a lima\c{c}on-shaped microcavity laser. Plots (a), (b), and (c) are marginal intensities as a function of $y$-coordinates at each deformation $\chi=0.43$, $\chi=0.454$, and $\chi=0.478$, respectively. They have maximal values around $y=\pm 0.99$ denoted by $y_{\max}$. Plots (d), (e), and (f) are conditional intensities as a function $x$-coordinate at the fixed value at $Y=y_{\max}$.}
\label{fig:2}
\end{figure*}
%%%%%

Figures~\ref{fig:1}(a), (b), and (c) display some profiles of the representative intensities of FFPs at each deformation $\chi=0.43$, $\chi=0.454$, and $\chi=0.478$ for the lima\c{c}on-shaped microcavity, respectively. For the case of an oval-shaped microcavity, Figures~\ref{fig:1}(d), (e), and (f), display some of the representative intensities of FFPs at each deformation $\varepsilon=0.043$, $\varepsilon=0.05$, and $\varepsilon=0.058$, respectively. The intensities are calculated from the transmitted rays using the Fresnel equations. To reproduce the previous results, we again consider transverse magnetic (TM) modes for a lima\c{c}on-shaped microcavity and transverse electric (TE) modes for an oval-shaped microcavity.

 In addition, to perform numerical calculations for the ray simulations, we considered the FFPs in the Cartesian coordinates within $-10\leq x\leq10$ and $-10\leq y\leq10$ for numerical convenience. That is, we discretized ($x$,$y$)-coordinates in the range of $x\in[-10,10]$ and $y\in[-10,10]$ into $2000\times2000$-grid. Additionally, the thick dotted lines in Figures~\ref{fig:1}(b) and (e) depict the peak intensities of directional emissions.

%%%%%
\section{Analysis on a lima\c{c}on-shaped microcavity} \label{sec:limacon}
%%%%%
\subsection{Marginal and conditional intensities in a lima\c{c}on-shaped microcavity} \label{subsec:limacon}
To address the peak emissions at the given FFPs in the Cartesian coordinate system, we first need to specify the locations of the $y$-coordinate for peak emissions because the directional light propagates along $x$-axis at a specific $y$-coordinate. For this approach, we introduce marginal intensity, which is defined as
\begin{align}
I(y)=\sum_{x_{i}=1}^{N} I(x_{i},y_{j})
\end{align}
with $N=2000$.
Plots (a), (b), and (c) in the upper panels of Figure~\ref{fig:2} present the results for each deformation parameter $\chi=0.43$, $\chi=0.454$, and $\chi=0.478$. We observe that the values of $I(y)$ have maximal values around $y=\pm0.99$ denoted by $y_{\max}$. Thus, we can also propose the quantity of the conditional intensity as a function $x$-coordinate $I(x)|_{y_{\max}}$ at the fixed value of $y_{\max}\simeq \pm 0.99$ to specify the FFPs for the peak emissions. The results are plotted in Figures (d), (e), and (f) in the lower panels of Figure~\ref{fig:2}. These plots can be interpreted as the peak intensities of directional light propagation in the microcavity lasers. Note that the overall values of $I(x)|_{y_{\max}}$ at $\chi=0.454$ are larger than those at $\chi=0.43$ and $\chi=0.478$. It is also expected that the relative portion $I(x\geq 0)|_{y_{\max}}/I(x\leq 0)|_{y_{\max}}$ at $\chi=0.454$ is larger than those at $\chi=0.43$ and $\chi=0.478$. Accordingly, this directly indicates that the forward emissions of FFPs at $\chi=0.454$ are larger than those of the others.

%%%%%
\subsection{Total and decomposed entropies in a lima\c{c}on-shaped microcavity}

%%%%%
\begin{figure}%[ht!]
\centering
\includegraphics[width=\linewidth]{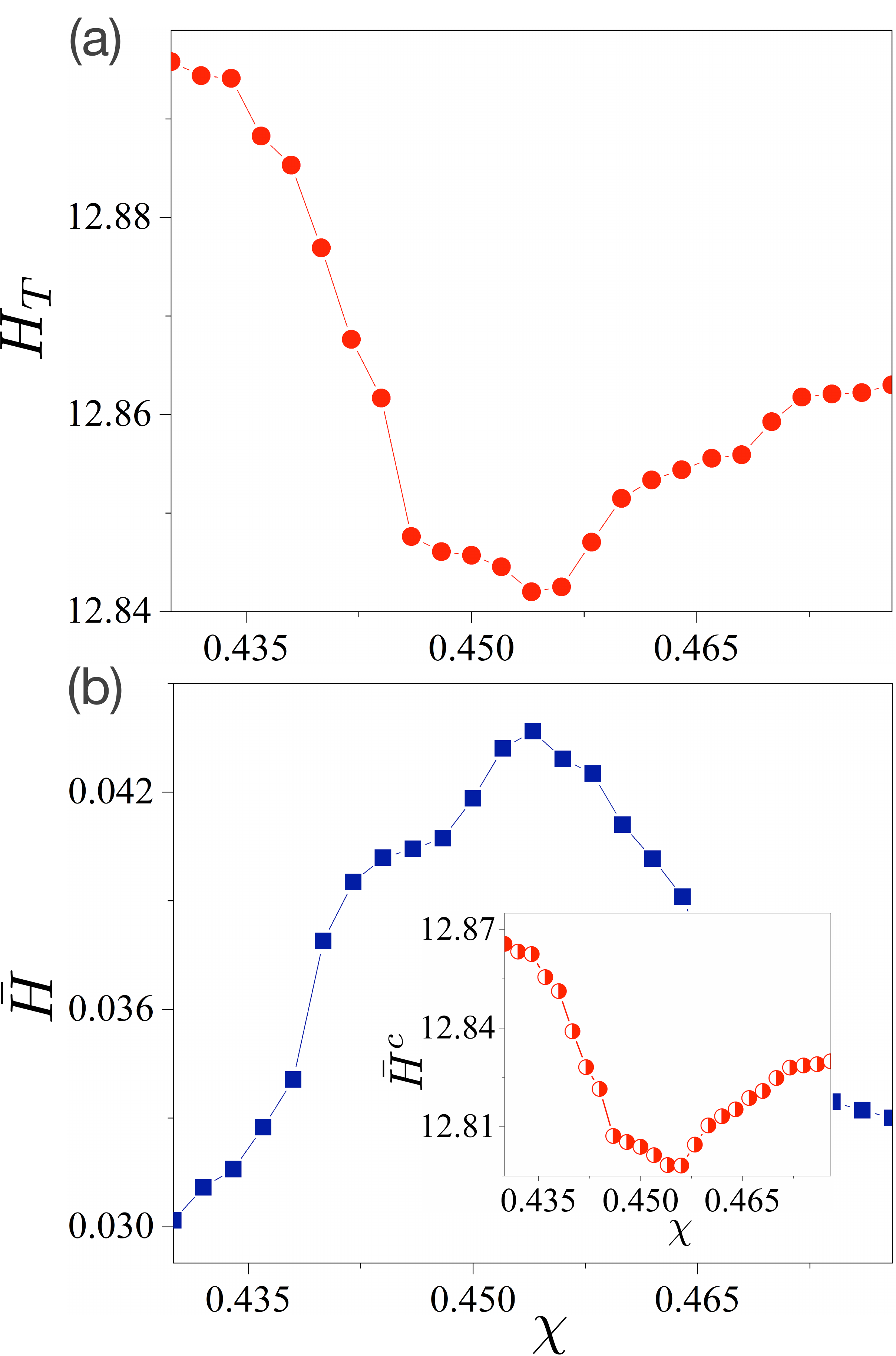}
\caption {Total and decomposed entropies in a lima\c{c}on-shaped microcavity laser. (a) Total entropy $H_{T}$ with joint probability $\rho(x,y)$ depending on the deformation parameter $\chi$. $H_{T}$ has a minimal value at $\chi=0.454$. (b) Decomposed entropy $\bar{H}$ of the peak intensities of FFPs. $\bar{H}$ has a maximal value at $\chi=0.454$. Inset in (b) denotes the complementary decomposed entropy $\bar{H}^{c}$.}
\label{fig:3}
\end{figure}
%%%%%

Directional emissions have been investigated in terms of directionality and divergence angle at given FFPs so far~\cite{MA17,JJ19,JM11,JA13,KC22}. In this sub-section, we present a method to estimate the degree of output power at given FFPs by exploiting the decomposed Shannon entropy. Accordingly, we should first have probability distributions of the given random variables associated with the FFPs. As mentioned in Subsec.~\ref{subsec:decomp} in Sec.~\ref{sec:decompent}, we consider the joint probability distributions $\rho(x,y)$ of the two random variables $X,Y$ by normalizing the intensity $I(x,y)$ of the FFPs under the condition $\sum_{i=1}^{N}\sum_{j=1}^{N} I(x_{i},y_{j})=1$ with $N=2000$. In this case, the total sample space $\Omega$ comprises $2000\times2000$-grid for the discretized ($x$,$y$)-coordinates in the range of $x\in[-10,10]$ and $y\in[-10,10]$.

Let us consider the total-sample space $\Omega$ related to the FFPs and analyze it using the total entropy $H_{T}$, which is defined as follows:
\begin{align}
H_{T}=-\sum^{N}_{i=1}\sum^{N}_{j=1}\rho({x_{i},y_{j}})\log \rho({x_{i},y_{j}}),
\end{align}
with $N=2000$. As shown in Figure~\ref{fig:3}(a), the value of $H_{T}$ is minimal at $\chi=0.454$, and its overall behavior is similar to the Shannon entropy of a single random variable angle $\theta$. (See details in our previous paper~\cite{KC22}.) This result indicates that the delocalization or complexity of the intensity distributions of the FFPs in the $(x,y)$-coordinates are minimized at $\chi=0.454$. Equivalently, the directionality of FFPs is maximized at $\chi=0.454$.

Now, we decompose the total-sample space $\Omega$ into sub-sample space $\bar{\Omega}$ and its complement $\bar{\Omega}^{c}$. Because defining the sub-sample space $\bar{\Omega}$ has no restrictions, we can select the normalized peak intensities of the directional emissions as the elements of $\bar{\Omega}$, which are depicted as dotted arrows in Figures~\ref{fig:1}(b) and (e), i.e., $\{\mathbb{P}(X\geq0, Y=y_{\max})\}\in\bar{\Omega}$. Note that we choose only $X\geq0$ to handle forward emissions solely. Thus, decomposed entropy $\bar{H}$ is defined as follows:
\begin{align}
\bar{H}=-\sum^{N/2}_{i=1}\rho({x_{i},y_{\max}})\log \rho({x_{i},y_{\max}}).
\end{align}

The explicit calculations are presented in Figure~\ref{fig:3}(b). Evidently, its absolute values are significantly smaller than those of $H_{T}$ in Figure~\ref{fig:3}(a). Most importantly, however, it should be noted that its overall pattern is opposite to the total entropy $H_{T}$ in Figure~\ref{fig:3}(a). The value of $\bar{H}$ is maximized at $\chi=0.454$. Therefore, we conjecture that $\bar{H}$ can estimate the output power because $\bar{H}$ is related to the spreading (or delocalization) of the intensity of the directional light propagation embedded in the FFPs. The inset of Figure~\ref{fig:3}(b) is complementary to $\bar{H}$. It behaves similarly to $H_{T}$, but is slightly less than the absolute value of $H_{T}$.

%%%%%
\section{Analysis on an oval-shaped microcavity} \label{sec:oval}
%%%%%
\subsection{Marginal and conditional intensities in an oval-shaped microcavity}
\begin{figure*}%[ht!]
\centering
\includegraphics[width=\linewidth]{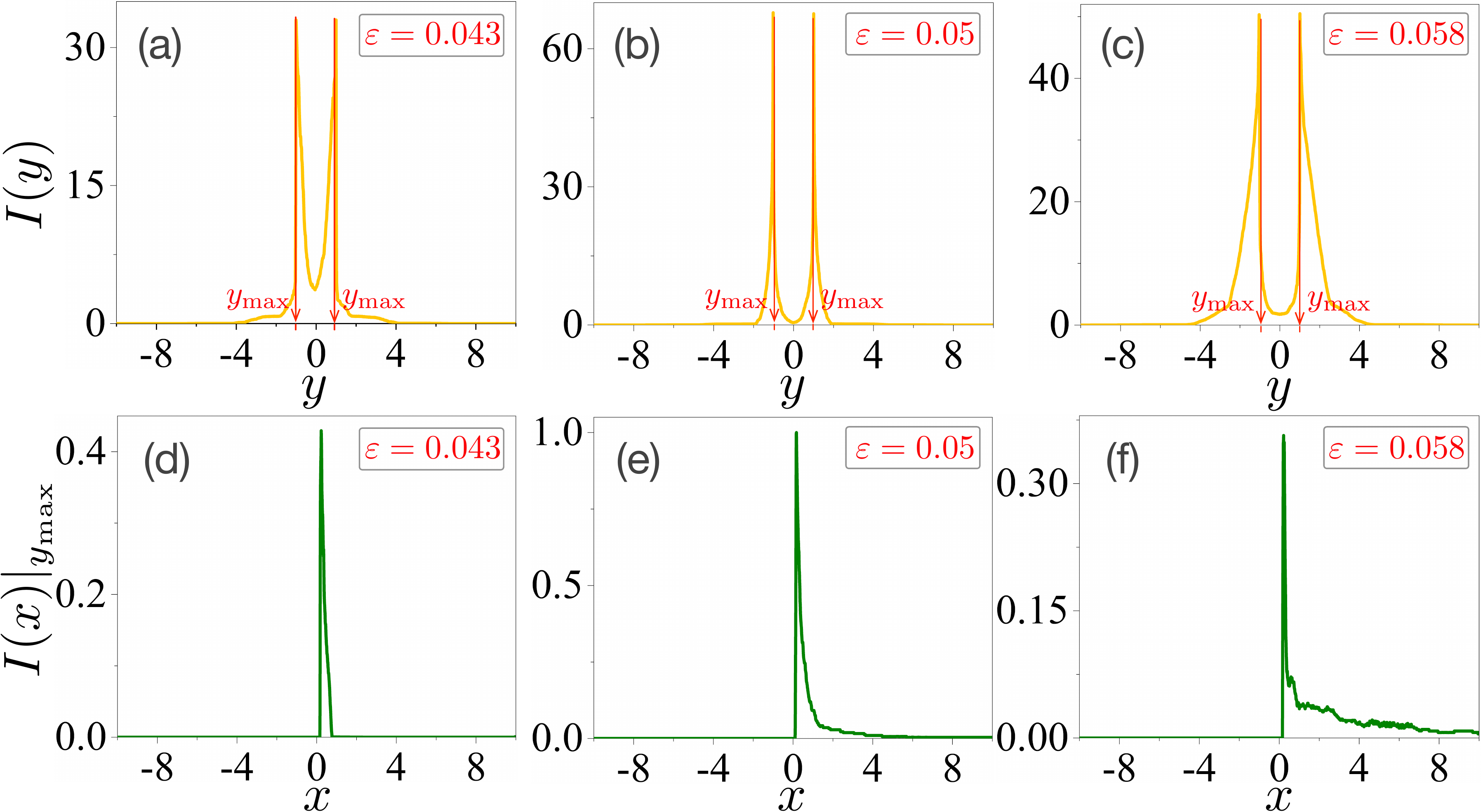}
\caption {Marginal and conditional intensities in the oval-shaped microcavity laser. Plots (a), (b), and (c) are marginal intensities as a function of the $y$-coordinate at each deformation $\varepsilon=0.043$, $\varepsilon=0.05$, and $\varepsilon=0.058$, respectively. They have maximal values around $y=\pm1.01$, as denoted by $y_{\max}$. Plots (d), (e), and (f) are conditional intensities as a function of the $x$-coordinate at the fixed value of $Y=y_{\max}$.}
\label{fig:4}
\end{figure*}

To validate the generality of our argument, we investigated the decomposed entropy of an oval-shaped microcavity laser. The marginal and conditional intensities of the FFPs in the oval-shaped microcavity laser is obtained in the same manner as in Subsec.~\ref{subsec:limacon} in Sec.~\ref{sec:limacon}, and Figure~\ref{fig:4} shows the results.

Figures~\ref{fig:4}(a), (b), and (c) present the marginal intensities as a function of the $y$-coordinate at each deformation $\chi=0.043$, $\chi=0.05$, and $\chi=0.058$, respectively. The values of $I(y)$ have maximum values around $y=\pm1.01$, as denoted by $y_{\max}$. The conditional intensities as a function of the $x$-coordinate at a fixed value of $Y=y_{\max}$ are plotted in Figures~\ref{fig:4}(d), (e), and (f). We also expect that the relative portions $I(x\geq 0)|_{y_{\max}}/I(x\leq 0)|_{y_{\max}}$ in Figures~\ref{fig:4}(d), (e), and (f) are larger than those shown in Figures~\ref{fig:2}(d), (e), and (f).  This can reveal that the forward emissions of the FFPs in the oval-shaped microcavity are larger than those in the lima\c{c}on-shaped microcavity.

%%%%%
%%%%%

%%%%%
\subsection{Total and decomposed entropies in the oval-shaped microcavity}
\begin{figure}%[ht!]
\centering
\includegraphics[width=\linewidth]{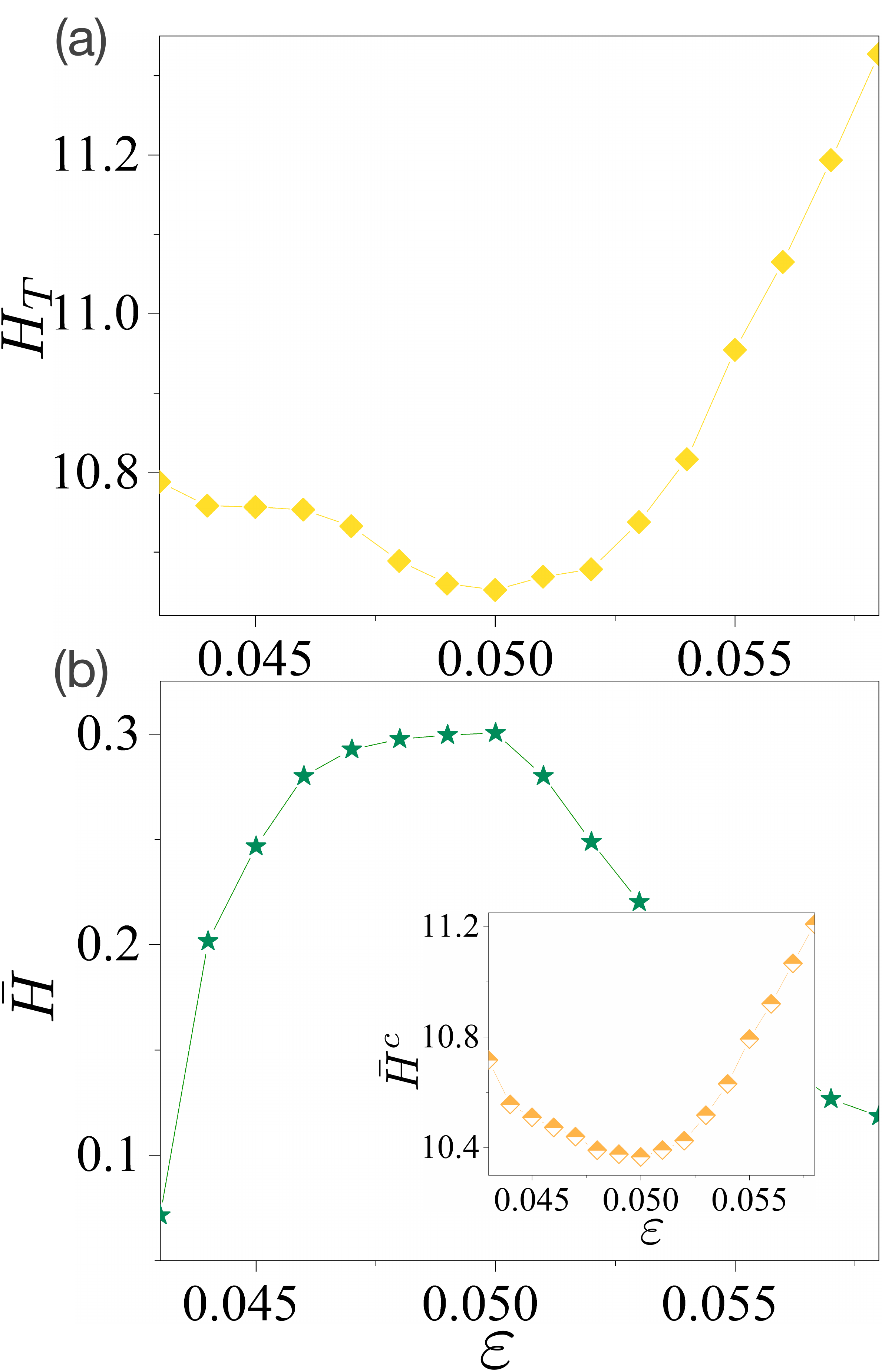}
\caption {Total and decomposed entropies in an oval-shaped cavity. (a) Total entropy $H_{T}$ with joint probability $\rho(x,y)$ depending on the deformation parameter $\varepsilon$. $H_{T}$ has a minimal value at $\varepsilon=0.050$. (b) Decomposed entropy $\bar{H}$ of the peak intensities of the FFPs. The entropy $\bar{H}$ has a maximal value at $\varepsilon=0.050$. Inset in (b) is the complementary decomposed entropy $\bar{H}^{c}$.}
\label{fig:5}
\end{figure}

The total entropy $H_{T}$ and decomposed entropy $\bar{H}$ are also obtained in the same manner as in Sec.~\ref{sec:limacon}. The results are presented in Figure~\ref{fig:5}. The value of $H_{T}$ in Figure~\ref{fig:5}(a) is minimized at $\chi=0.05$, and its overall values are smaller than those in Figure~\ref{fig:3}(a). These results agree with our previous results~\cite{KC22}. However, unlike the case of the lima\c{c}on-shaped microcavity, its overall behavior exhibits a slight difference from that of the Shannon entropy of a single random variable angle $\theta$. The value of $H_{T}$ at $\varepsilon=0.058$ is larger than that of $H_{T}$ at $\varepsilon=0.043$, whereas the Shannon entropy in Ref.~\cite{KC22} shows the opposite behavior. This can be attributed to the fact that the primary emission route of the FFPs in the lima\c{c}on-shaped microcavity hardly changes depending on the parameter $\chi$, whereas that of the FFPs in the oval-shaped microcavity changes depending on $\varepsilon$. That is, the FFPs in the oval-shaped microcavity had four primary emission routes at $\varepsilon=0.043$, whereas they had six primary emission routes at $\varepsilon=0.058$. Therefore, the total Shannon entropy of the probability distributions of a single random variable $\theta$ cannot capture the total information of the FFPs, whereas those of the two random variables ($X,Y$) can.

%%%%%
%%%%%

The decomposed entropy $\bar{H}$ of the peak intensities of the directional emissions is shown in Figure~\ref{fig:5}(b). It is maximized at $\chi=0.05$ with a value of $\bar{H}\simeq 0.3$. Thus, we can suggest that the main results are as follows. The degree of directional emission defined by the intensity of angle $I(\theta)$ shows a similar order of magnitude for the two different microcavity lasers, that is, directionality defined by $U_{W}$, $U_{C}$, or $S_{N}$, and the divergence angle exhibits a similar order of magnitude in both microcavity lasers in Ref.~\cite{KC22}. However, the decomposed entropy $\bar{H}$ in Figure~\ref{fig:5}(b) is almost ten times larger than that of the decomposed entropy $\bar{H}$ in Figure~\ref{fig:3}(b). Accordingly, the decomposed entropy $\bar{H}$ can more sensitively capture the properties of directional emissions for different microcavity lasers. Here, we also conjecture that the decomposed entropy $\bar{H}$ can estimate the output power at the given FFPs without any further information.

%%%%%
\section{Conclusion} \label{sec:conclusion}
We introduced the total entropy and decomposed entropy for the joint probabilities of two random variables $(X,Y)$ and examined the possibility of estimating the output power of deformed microcavity lasers at the given FFPs. The total entropy $H_{T}$ is defined in the total sample space $\Omega$, whereas the decomposition entropy $\bar{H}$ is defined in the sub-sample space $\bar{\Omega}$, which consists of normalized peak intensities of the directional emissions. The total entropy $H_{T}$ behaves roughly similar to the Shannon entropy for a single random variable $\theta$. On the contrary, the decomposed entropy $\bar{H}$ behaves in the opposite way to $H_{T}$. The decomposed entropy $\bar{H}$ in the oval-shaped microcavity is ten times larger than the decomposed entropy $\bar{H}$ in the lima\c{c}on-shaped microcavity. Thus, $\bar{H}$ can more efficiently detect the properties of directional emission for different microcavity lasers than directionality and divergence angles. It can also estimate the degree of the output power at given FFPs without any further information. The output power plays a crucial role in designing microcavity lasers as quality factor and  directional emission do. In addition, because our notion depends only on the given FFPs, it can be applied to any type of microcavity laser or antenna system. Accordingly, we hope that our results can help in designing high-performance lasers or other physical systems concerning directional emissions.

Finally, it is worth mentioning that the decomposed entropy is quite different from the conditional entropy; more precisely, the entropy of $X$ conditioned on $Y$ taking the value $Y=y_{\max}$, i.e., $H(X|Y=y_{\max})$. In this case, the total sample space $\Omega$ collapses into the sub-sample space $\bar{\Omega}$ and loses information about its complement $\bar{\Omega}^{c}$.

%%%%%
\section*{Acknowledgments}
This work was supported by the National Research Foundation of Korea and a grant funded by the Ministry of Education (Grant Nos. NRF-2021R1I1A1A01052331 \& NRF- 2021R1I1A1A01042199) and the Ministry of Science and ICT (Grant No. NRF-2020M3E4A1077861).

%%%%%
%

\end{document}